# Coherent beam combination of ultrafast fiber lasers

Arno Klenke, Michael Müller, Henning Stark, Marco Kienel, Cesar Jauregui, Andreas Tünnermann and Jens Limpert

*Abstract*—The performance of fiber laser systems has drastically increased over recent decades, which has opened up new industrial and scientific applications for this technology. However, currently a number of physical effects prevents further power scaling. Coherent combination of beams from multiple emitters has been established as a power scaling technique beyond these limitations. It is possible to increase the average power and, for pulsed laser systems, also parameters such as the pulse energy and the peak power. To realize such laser systems, various aspects have to be taken into account which include beam combination elements, stabilization systems and the output parameters of the individual amplifiers. After an introduction to the topic, various ways of implementing coherent beam combination for ultrashort pulses are explored. Besides the spatial combination of beams, the combination of pulses in time will also be discussed. Recent experimental results will be presented, including multi-dimensional (i.e. spatial and temporal) combination. Finally, an outlook on possible further developments is given, focused on scaling the number of combinable beams and pulses.

*Index Terms*— Coherent combining, Fiber lasers and amplifiers, Ultrafast optics

## I. INTRODUCTION

High repetition rate ultrafast laser systems have established themselves as an enabling technology for a wide variety of industrial and scientific applications, ranging from materials processing to exploring high-field physics. These applications are currently driving the demand for ever increasing performance figures. Coherent beam combination is one approach to meet these requirements. The basic idea is actually very simple: Use multiple laser emitters and combine their outputs to achieve performance levels otherwise unattainable with a single emitter. The particular effect posing the strongest limitation in single emitter systems depend on the considered performance figure, e.g. average power, pulse energy, peak power, and the laser architecture.

Until now, coherent combination has been mostly applied to fiber laser systems. Therefore, this contribution focuses on the specific properties and limitations of fiber laser systems, although analogous discussions could be made for other laser architectures. These include bulk lasers [1], slab lasers [2], semiconductor lasers [3] or optical parametric amplifiers [4]. While fiber laser systems have been known for producing high average, the transverse mode instability effect [5]–[7] currently limits the average output power of diffraction-limited beams to 1 kW [8] in femtosecond pulse operation (see Fig. 1).

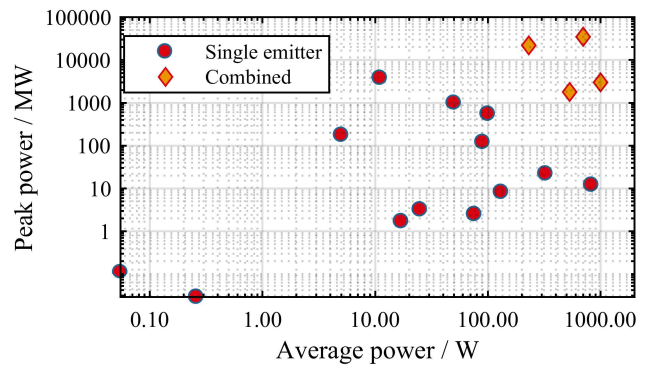

Fig. 1. Peak and average power values of selected single emitter femtosecond fiber laser systems (red circles) and recent results with coherent beam and pulse combination (orange diamonds).

Although different mitigation strategies have been developed, this problem remains the focus of current research. In the case of pulse energy and peak power scaling, non-linear effects [9] and, ultimately, self-focusing [10] in the fiber set the upper limit. Micro-structured fibers allow increasing the mode-field diameter in order to mitigate non-linear effects, but this approach has its own technical limitations. The same holds true for the chirped-pulse amplification (CPA) [11] concept, where the stretched pulse duration is ultimately limited by the grating sizes in the pulse compressor. As can be seen in Fig. 1, the peak power of femtosecond pulses emitted by CPA fiber laser systems remains below a value of 10 GW [12]. Despite possible further performance scaling of fiber laser systems, it is not expected that a single fiber will be able to provide drastically higher performance in the near future. Hence, coherent

This paragraph of the first footnote will contain the date on which you submitted your paper for review. This work has been partly supported by the Free State of Thuringia (2015FE9158) "PARALLAS" and the European Research Council under the grants agreements no. 670557 "MIMAS" and no. 617173 "ACOPS".

All authors are or have been with the Institute of Applied Physics, Abbe Center of Photonics, Friedrich-Schiller-Universität Jena, 07745 Jena, Germany (e-mail: a.klenke@gsi.de, michael.mm.mueller@uni-jena.de, lars.henning.stark@uni-jena.de, cesar.jauregui-misas@uni-jena.de, andreas.tuennermann@iof.fraunhofer.de, jens.limpert@uni-jena.de). Additionally, A.Klenke, J. Limpert and A. Tünnermann are with the Helmholtz-Institut Jena, 07743 Jena, Germany. J. Limpert and A. Tünnermann are also with the Fraunhofer Institute for Applied Optics and Precision Engineering, 07745 Jena, Germany. Marco Kienel is now with Active Fiber Systems GmbH, 07743 Jena, Germany (e-mail: kienel@afs-jena.de).

combination has been applied to such systems for quite some time now and has successfully pushed the performance beyond the current limitations of single emitter systems (see Fig. 1), which will be discussed in this paper.

Coherent combination was already demonstrated in 1970 with the combination of two semiconductor oscillators [13]. Today, coherent combination covers a wide field of different implementations, which makes a compact and comprehensive description challenging. The first differentiation that can be done is to separate it from non-coherent beam combination techniques. In the strict sense, the output beams from the various emitters have to be able to interfere which each other to be called coherent. This means that technologies such as spectral combination [14], where the emission at different wavelengths is combined, fall outside of the scope of coherent combination, even though these concepts often pursue similar performance scaling goals. The different emitters to be coherently combined can either be independent laser systems, coupled oscillators, amplifiers or also multiple temporally separated pulses emitted by a single amplifier. In the case of the coupled oscillators, the mutual coherence is achieved through the coupling process. For the combination of amplifiers, they can be seeded by a common frontend, thus creating the precondition of coherence between the output beams after amplification.

This paper focuses on the spatial combination of amplifiers and, additionally, on the temporal pulse combination, because this architecture is the most suitable for high power systems. Using both techniques in a single laser system is known as multidimensional coherent combination.

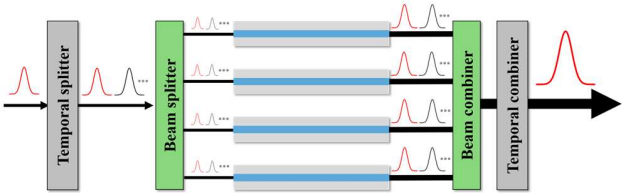

Fig. 2. Schematic setup of coherent beam combination of multiple amplifiers. Additional temporal splitting and combination also shown to realize multidimensional pulse combination.

The corresponding basic schematic setup is shown in Fig. 2. The input from a frontend is first split into multiple amplifier channels using a beam splitter. Then, after independent amplification of each channel, the different output beams of the channels are recombined using a beam combiner. Independently of the actual implementation of the beam combiner, the beams will have to be phased correctly to interfere constructively. In the case of the combination of continuous-wave radiation a control of the phase in the range of $2\pi$ (i.e. sub-wavelength phase control) is often sufficient because the coherence length of the output beams is longer than any potential path-length mismatches between the amplifier channels. However, in the case of pulsed operation, the coherence length of the light signal is significantly shorter, which demands precisely matched pulse propagation distances in addition to the mentioned sub-wavelength phase control [15]. While there are coherent combination schemes that lead to an intrinsic matching of the path-lengths [16], these schemes are generally not scalable to a higher number of channels, and, therefore, active stabilization is required because during operation, thermal and environmental drifts in the different channels are unavoidable. Hence, a phase difference detection technique has to be implemented that drives the beam phasing elements in each channel. In general, this spatial combination technique can be thought of as an artificial increase of the mode-field diameter of the fiber, without the associated detrimental effects, such as a reduced beam quality. Temporal pulse combination, often called divided-pulse amplification (DPA), can be additionally or exclusively implemented in such a setup incorporating a temporal pulse splitter to create multiple temporal pulse replicas. After amplification, a corresponding combiner for the subsequent pulse recombination [17], [18] should also be introduced in the system. However, some approaches employ the intrinsic repetition rate of the frontend and additional phase modifications, instead of an additional temporal pulse splitter [19], [20]. Consequently, divided-pulse amplification is analogue to an artificial increase of the temporal pulse duration and, therefore, it can lead to an increase of the achievable pulse energy and peak power.

In the following, after some general considerations, the three major components of spatial beam combination will be discussed: the amplifiers, the beam splitters and combiners and, finally, path-length stabilization mechanisms. This is followed by the description of temporal pulse combination. Compared to previous publications reviewing the state of this field [21], novel results for femtosecond fiber lasers, including high-power multi-dimensional combination, will be presented. Additionally, an outlook on future prospects of coherent combination is given, focused on integrated approaches for achieving high channel counts.

## II. GENERAL CONSIDERATIONS

The most important performance figure for the combining processes is its power efficiency $\eta_{\text{comb}}$. It comprises both the spatial combination efficiency $\eta_{\text{spatial}}$ and the temporal combination efficiency $\eta_{\text{temporal}}$ depending on the employed combination scheme. On the one hand, $\eta_{\text{spatial}}$ is defined as the ratio of the power in the combined beam $P_{\text{comb}}$ to the sum of all power $P_i$ emitted by the individual channels. On the other hand, $\eta_{\text{temporal}}$ is the percentage of the total output energy $E_{\text{total}}$ after spatial combination that is temporally recombined into the single output pulse with energy $E_{\text{pulse}}$:

$$\eta_{\text{comb}} = \eta_{\text{spatial}} \eta_{\text{temporal}} = \frac{P_{\text{comb}}}{\sum_{i}^{n} P_i} \frac{E_{\text{pulse}}}{E_{\text{total}}}$$

In the best case, a value of 1 is achievable and, therefore, a result as close as possible to 1 is desired. Hereby, smaller differences between the channels lead to a higher achievable combination efficiency [15], [22], [23]. A high combination efficiency also has consequences for the combined emission. The higher this value, the closer the properties of the combined emission (e.g.

intensity and phase profile) will be to the ones of the individual channels. Hence, laser systems based on coherent combination mostly preserve parameters such as beam quality and pulse quality from the single amplifier systems. Moreover, some channel dependent spatial and temporal artifacts can actually be reduced by the combination. Thus, it is possible that parameters such as the beam quality are improved for the combined beam compared to the single channels. Another interesting aspect is the evolution of the combination efficiency with the number of channels. If one assumes independent amplification as shown in Fig. 2, meaning that no intra-channel crosstalk takes place, the efficiency is, from a theoretical point of view, expected to converge to a fixed value. Please note that this is only valid if there are no losses in the beam combiner that depend on the channel count. Hence, the technical realization of systems with increasingly higher number of amplifier channels, while controlling the complexity and cost of such systems, remains a focus of development.

### III. Amplifiers for Coherent Combination

The coherent combination concept is independent from the applied amplifier architecture. As long as multiple amplifiers emit mutually coherent beams, it can be implemented. However, fiber lasers are mostly used due to their simple single-pass design. Additionally, they emit a high quality and reproducible beam, allowing for a high combination efficiency. The impact of the amplifiers on the temporal intensity and phase profile of the pulses also has to be considered. Due to dispersion (e.g. propagation through material), saturation or non-linear effects such as SPM, these properties can change. This is not an issue as long as the pulses in the different channels are affected in a similar way. However, channel differences (e.g. fiber length differences, coupling loss differences) have a detrimental influence on the combination efficiency and have to be kept low [15].

### IV. Beam Splitters and Combiners

For a first rough classification, the various coherent beam combination implementations can be divided in two different approaches: tiled-aperture and the filled-aperture. In the first one, a larger beam is constructed from many beams placed adjacently to each other, e.g. in a hexagonal beam array. Phase control of the individual beams is applied to achieve a flat spatial phase in the combined beam. Natural divergence causes the beams to overlap in the far field after propagation [24]. An advantage of tiled combination is that it requires no beam combination element, which might be prone to power scaling limitations. A disadvantage, however, is the technical challenge related to minimizing the gap between the sub-beams when they are arranged next to each other. In most cases, each of the sub-beams has a Gaussian-like beam profile, which means that the combined beam will have an intensity modulation across its cross-section. This leads to significant power loss into side lobes in the far field, resulting in a reduced achievable combination efficiency.

The other well-known method for beam combination is the filled-aperture approach. Here, the different sub-beams are superimposed both in the near- and in the far-field. If one assumes identical spatial phase and beam profiles, the resulting combined beam will be indistinguishable from any of the sub-beams and, thus, the beam quality is maintained. This approach requires combination elements to actually achieve the beam overlap. The simplest implementations are intensity beam combiners, which are partially reflective and transmissive.

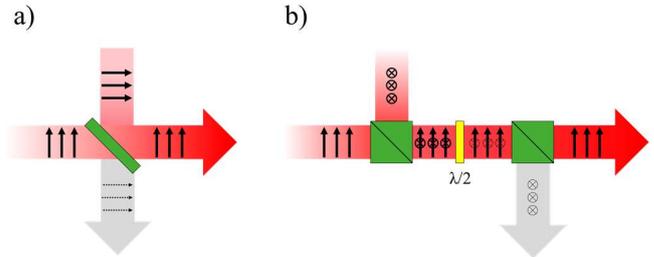

Fig. 3. Combination of 2 beams using a) a 50% intensity beam combiner with p-polarized input beams and b) a polarization beam-splitter (PBS) with additional half-wave-plate and analyzer to let the 2 beams interfere. The arrows denote the polarization state of the beams and the dashed lines show the non-combining parts.

If one considers such an element with 50% reflectivity, the combined power of two identical incident beams (e.g. both with p-polarization) can be emitted in any arbitrary ratio from the two output ports depending on the phase difference between the incident beams, as shown in Fig. 3a). Hence, such an element can be used for beam combination if the phases of the beams are controlled. A more flexible implementation is based on polarization beam-splitter/combiners (PBS) [25] followed by a polarization analyzer (Fig. 3b)). Here, an s- and a p-polarized beam of arbitrary power ratio are superposed into a single beam. The output state of polarization depends on the relative phase between the two input beams. If the phase difference is zero, the polarization state is linear and can be rotated to p-polarization using a half-wave plate. Thus, the complete power can be passed through a subsequent analyzer. The advantage of polarization beam combining is its flexibility with respect to the input beam power ratio.

Intensity and polarization beam combiners inherently combine two beams at a time. A larger number of beams can be combined by cascading these elements. A system using this approach with up to 8 channels will be presented later in this paper. However, the required component count will grow roughly linearly with the number of beams. Therefore, non-cascading N:1 beam combiners are of great interest.

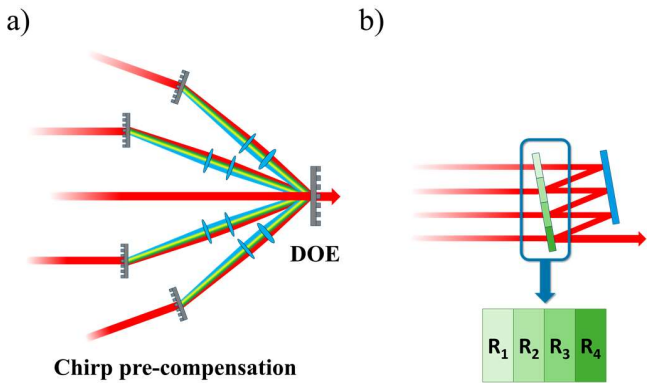

Fig. 4. Beam combination using a) a DOE with broadband pulses and spatial chirp pre-compensation with additional gratings and imaging optics and b) a SMS component consisting on a highly reflective mirror and a partial reflector with zones of different reflectivity.

A common concept is the use of diffractive-optical elements (DOE) [26]. Those have already been successfully applied to high power CW systems for the combination of up to 15 beams [27]. The application to wide bandwidth sources for short-pulse operation is, however, more challenging. The combined beam will have a spatial frequency chirp due to the diffraction. This, in turn, will lead to a loss of combining efficiency. Therefore, a channel dependent pre-compensation of this chirp is necessary, which has recently been demonstrated by using a second array-DOE [28].

Another integrated array approach for intensity beam combining is a segmented mirror splitter (SMS). This optical element was first described as a splitter for temporal pulse combination [29], but can be used as a spatial beam-combiner as well. The structure of a SMS is shown in Fig. 4b). It consists of two parallel optical elements, one highly reflective mirror and one partial reflector with zones of different reflectivity. The reflectivities for incident beams with the same power can be calculated by the formula $R_x=1-(1/x)$, where x is the beam number between 1 and N. The first beam sees no reflectivity (i.e. this first zone is anti-reflection coated) and interferes with the second beam on its return to the partial reflector, with a reflectivity of 50% in this zone. The combined beam continues the zig-zag path and gains more power at every point of interference, with each zone of reflectivity matched to the power ratio. This setup is able to handle high-average-powers, as the combined beam is always reflected. The working principle is very similar to passive enhancement cavities, where average powers up to the MW-level have been demonstrated [30]. Additionally, the impact of diffractive effects is negligible, thus this component is also suitable for wide-bandwidth sources. However, due to the multiple reflections of the beam, the employed optical elements need to be of very good optical quality and surface deformations have to be avoided. By using a second SMS in the orthogonal direction, the combination can be extended from a one-dimensional beam array to a two-dimensional array. As can be deduced from Fig. 4b), the incident beams have to arrive at the component with fixed temporal delays determined by the distance between the two optical elements. This temporal delay can be created by using the SMS setup in the opposite direction as a beam splitter before amplification. In fact, all the previously described beam combiners can be used in this configuration. The choice of the appropriate combiner depends on the number of beams to be combined, the average power and spectral bandwidths of the system.

## V. Beam Phasing

The last component necessary for most coherent beam combination setups is an active phase stabilization consisting of a detection of the phase differences between the individual beams and phase actuators to compensate for them. A variety of implementations exists for each beam-combiner architecture. For polarization beam combination, Hänsch-Couillaud [31] (HC) detection is an established technology to measure the phase difference of two orthogonally polarized beams. For intensity beam combination, methods based on applying small phase modulations, followed by demodulation of the combined signal at a final detector such as LOCSET [32] or SPGD [33] can be employed. There are other concepts for tiled-aperture combination that do not need to apply phase modulations on the signal [24]. Independent from the actual implementation, the error signal generated by the detection electronics has to cover the frequency range of the occurring phase fluctuations. For most fiber laser systems, a bandwidth of around 1 kHz [34] is usually sufficient. The choice of phase actuators often depends on the operation mode of the system. Phase actuators with a $2\pi$ phase range are sufficient for CW laser systems, whereas for pulsed laser systems, phase jumps during operation that lead to a decrease of the temporal pulse overlap should be avoided. Therefore, manual delay stages are often employed to coarsely match the path-lengths in the different channels to within a few wavelengths before using piezo based actuators for fine path-length matching. The advantage of piezo actuators is that they provide a sufficient piston range during operation. It should be noted that the mentioned mechanisms can only stabilize the system to a local maximum of constructive interference, but additional mechanisms for absolute path-length matching are being investigated [35]. In addition to path-length stabilization, the automatic alignment of the beams can also be realized similarly [36].

## VI. Temporal Pulse Combination

Coherent combination of laser pulses is not exclusively used in spatial domain, but it can also be applied to the temporal domain. Amplification and subsequent temporal combination of pulse sequences into a single energetic pulse is a proven approach to peak-power and pulse-energy scaling beyond CPA. This scheme is analogous to having a significantly increased effective pulse duration during the amplification. Thus, the peak intensity in the optical materials and, therewith, the impact of nonlinear effects are reduced. In the recent years, various techniques have been developed based on this idea.

First among these was a method which was initially introduced for excimer lasers [17] and which was later also applied to femtosecond pulses, named divided-pulse amplification (DPA) [18]. Thereby, a single laser pulse is temporally separated into a pulse burst by a set of optical delays. The burst is then amplified and, finally, stacked into a single pulse again. The recombination of the pulses is accomplished in the same setup used for the pulse splitting just by rotating the polarization and by reflecting the burst backwards through the setup. This implementation, being a Sagnac interferometer, is called passive DPA as it does not require phase stabilization [16]. When aiming for high extraction efficiency from the amplifiers, however, gain saturation leads to amplitude deformation of the pulse burst which reduces the combining efficiency. A possible solution is to separate the pulse division and recombination into two separate stages, allowing for gain saturation compensation. Since this type of DPA requires an active phase control to maintain constructive interference, it is known as actively controlled DPA [37]. However, the control over the single pulses is limited, leaving some residual phase differences and/or amplitude mismatches between the different pulse replicas which reduce the efficiency. Hence, as an improvement, electro-optically controlled DPA [19] has been proposed, which makes an extra pulse division setup redundant, which is mostly fiber-integrated and which allows for both phase and amplitude control of each pulse of the burst. This technique potentially allows keeping the overall efficiency high, while significantly improving the interferometric stability and reducing the setup complexity.

An alternative approach is the so-called coherent pulse stacking [38], which employs low-finesse Gires-Tournois interferometers (GTI) to stack long trains of pulses. These very low-finesse cavities are length matched to the repetition rate of the input pulse train. The pulses require matched amplitudes and phases to interfere constructively with the preceding pulses which have already been stacked in the GTI. The last pulse of the pulse train has a π phase shift, causing constructive interference on the outside of the GTI, thus coupling out the pulses circulating in the GTIs.

A third technique is based on a passive enhancement cavity, in which a train of similar pulses is stacked [39], similarly to the coherent pulse stacking technique. The out-coupling itself, however, is accomplished with a fast optical switch.

## VII. COHERENTLY COMBINED FEMTOSECOND FIBER LASER SYSTEMS

Coherent beam combination, especially of femtosecond fiber laser systems, has seen a rapid development, going from proof-of-principle experiments [40], [41] to record power levels beyond the capabilities of single amplifier systems in a just a few years.

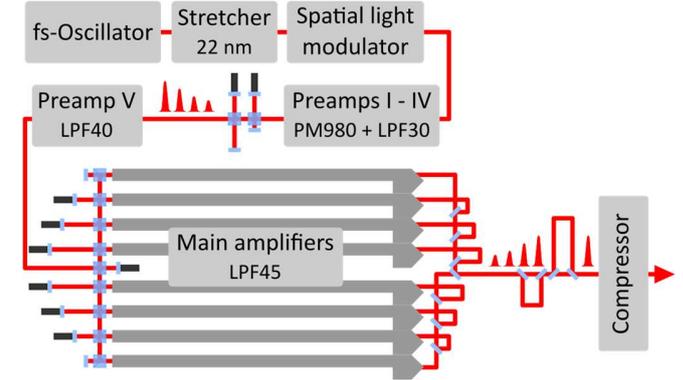

Fig. 5. Schematic setup of a femtosecond fiber laser system with a main amplification stage comprising 8 parallel channels. Additional temporal pulse splitting into 4 replicas and corresponding recombination stages can be added to realize multi-dimensional pulse combination.

Today, systems with up to 8 parallel amplifiers based on advanced large-area mode fibers (large-pitch fibers [42]) have been demonstrated. In this setup, as shown in Fig. 5, 1.1 m long rod-type fibers with mode-field diameters of 64 μm each are used. The splitting and combination is achieved using polarization beam-splitters and combiners in a cascaded arrangement. A HC stabilization system is implemented by measuring the polarization state after every combination step. The generated error signals are then fed back through PID regulators to piezo-mounted mirrors to adjust the path-lengths accordingly. A frontend consisting of a femtosecond oscillator followed by a preamplifier chain is used as the seed system. It supports flexible repetition rates by means of two acousto-optic modulators (AOM) and emits stretched femtosecond pulses (around 1.3 ns stretched pulse duration) at 27 W of average power. While polarization beam-splitter cubes are used in the splitting stage, the beam combination comprises thin-film polarizers (TFP) which are able to support the high output average powers of up to 145 W per channel. After combination, the beam is guided through a grating compressor [43] to restore the femtosecond pulse duration at the output. In a first experiment, 1 kW average power at 1 MHz repetition-rate has been achieved with 260 fs pulse duration [44]. A very high combination efficiency of 91% was measured, matching well the theoretical expectations. Additionally, the beam quality is a near diffraction-limited value of $M^2 < 1.1$.

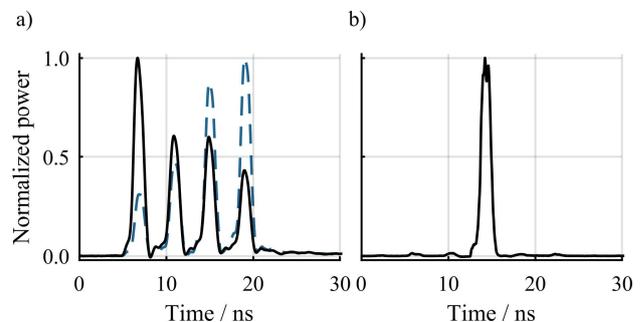

Fig. 6. a) The 4 temporal pulse replicas before amplification (dashed blue), after amplification (black) and b) the temporally combined pulse

While it was possible to increase the pulse energy in the same configuration, additional temporal pulse combination is implemented to see a drastic improvement of this value. Each input pulse of the seed system is split up using two free-space delay lines based on PBSs to generate 4 pulse replicas with a spacing of 4 ns each. This spacing is required so that the stretched pulses do not overlap temporally, which might cause pulse distortions in the amplifiers due to cross-phase modulation or four-wave mixing. A similar delay stage is placed between the beam combination stage and the compressor (again based on TFPs). The pulse train is shown in Fig. 6, before amplification, after amplification and after the temporal recombination. As can be seen, the input pulse train has been shaped to compensate for the saturation of the amplifiers. However, contrary to expectations, it is not set up to result in a flat output pulse train. Instead, the shaping was optimized for the best combination efficiency, which is achieved when the B-integrals of the pulse replicas are matched [45].

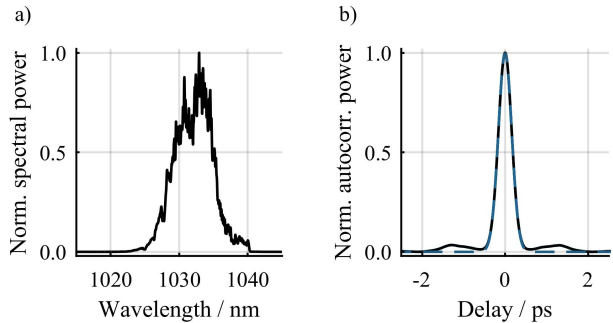

Fig. 7. A) Combined spectrum and b) autocorrelation trace (black) and transform-limited autocorrelation trace (dashed blue) for the 8 channel laser system with 4 temporal replicas.

Using this method, the spatial and temporal combined and compressed pulse energy could be increased to 12 mJ, while sustaining an average power of 700 W [45]. A pulse duration of around 270 fs was estimated from the autocorrelation trace in Fig. 7b) and the high beam quality emitted by the individual fiber channels is maintained in the combined beam. LOCSET phase stabilization was applied for the DPA as the pulses in the burst had alternating perpendicular polarization states, which prevents the use of HC in the standard configuration. Furthermore, LOCSET was modified using a temporal gating of the input signals to frame just the time window in which the correctly combined pulse is expected. Otherwise, the temporal combination of multiple pulses has intrinsically several local maxima of the combination efficiency where the power is still distributed over multiple temporal pulse replica [46]. While a similar modification is possible for HC, the temporal gating configuration would be more challenging in this case because multiple temporal gates per pulse train would be required. As shown in Fig. 1, the achieved peak power of 35 GW in combination with the already mentioned 700 W average power drastically surpasses the performance of single amplifier systems. The result demonstrates the advantages of the coherent combination concept.

At this point it should be mentioned that one of the most prominent challenges for the realization of such systems has been the handling of the resulting average and peak power. While the latter one [47] can be compensated for by increasing the beam diameter and by adapting the system design, the first one is harder to solve. The components handling the high power beams, such as the beam-combiners themselves or lenses, need to be carefully selected in order to avoid thermal effects. Those not only have a detrimental effect on the output beam quality, but can also reduce the overall combination efficiency. Hence, for systems moving to the multi-kW average power range, the use of reflective optics will become more and more important. The best way to prove the viability of a laser concept is to put it into use for the targeted applications. It has already been shown that non-linear pulse compression [48] of coherently combined fiber laser systems results in an excellent driving source for high-harmonic generation with a photon-flux in the XUV wavelength region that can often challenge large-scale synchrotrons [49]. The generated XUV radiation has already enabled applications such as the investigation of molecule dynamics [50] or nanoscale diffractive imaging with high temporal resolution [51]. With a future performance scaling, these applications can reach a new quality, e.g. doing coherent diffractive imaging or tomography of samples with video framerate.

## VIII. FUTURE SCALING PROSPECTS

The most prominent question in coherent beam combination is about the scalability to a high channel count. While it is possible, from a theoretical point of view, the actual experimental realization is challenging, which is why most systems demonstrated to date still have a rather low channel count. However, technologies supporting a high channel count are currently in development, including the already mentioned N:1 beam combiners and scalable phase detection systems. Currently, the components in almost all setups are identical to the ones used in single amplifier systems. Thus, the component count grows linearly with the number of channels.

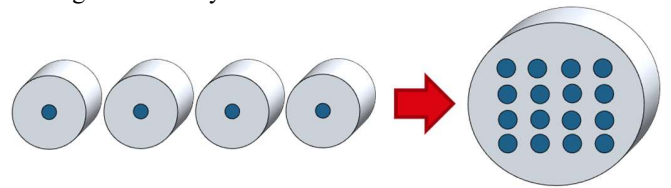

Fig. 8. Schematic presentation of the integration of multiple fibers into a single multicore fiber with a common pump cladding for all cores.

Work has already been started on integrating multiple signal cores into a single fiber, as shown in Fig. 8, and first proof-of-principle experiments regarding their signal combination have been carried out [52]–[54]. The scaling of the total mode-field diameter (adding all cores) in the same way as for multiple single-core amplifiers is straightforward and will increase the pulse energy and peak power handling capabilities equally. The average power scaling of these structures is the focus of ongoing research. First results indicate a linear improvement of the mode-instability threshold with the number of cores, deduced from an experiment with 4 cores in a row [55]. Additionally, very high average powers (far beyond the mode-instability threshold) have already been extracted from an all-fused-silica rod-type LPF fiber [56], proving the basic power handling capabilities of such fibers. The area of a shared pump cladding grows linearly with the integrated cores and, therefore, sufficient pump power is available from semiconductor pump diodes by keeping the required brightness value constant with the higher power demand.

Another major point of research in the near future will be the development of compact multichannel phase actuators. This includes the integration of known technology such as piezo actuators, EOMs or AOMs into compact array packages. Additionally, new microelectromechanical [57] might be applicable here. Major concerns will be cost, control speed and the required piston range of at least a few µm for short-pulse operation.

For the temporal pulse combination, the scaling prospects are less straightforward. Ultimately, the stored energy in the fiber sets an upper limit to the achievable pulse energy. The number of temporal replicas required to extract most of this energy depends on the pulse duration and, therefore, on the used stretcher and compressor in CPA systems. In order to simplify this concept, as already mentioned in section VI, the pulse splitting stage can be replaced by a setup that modifies the pulse train at the fundamental repetition rate emitted by the frontend system. Additionally, the multi-nanosecond delay lines required for temporal combining after amplification can be made more compact by using Herriott cells [58], which can also avoid propagation induced beam parameter mismatches for the different temporal delays due to their imaging capabilities.

| Pulse | 1 | 2 | 4 | 8 | 16 | 32 | Stored |
|---|---|---|---|---|---|---|---|
| Energy | 900 µJ | 1,5 mJ | 2,4 mJ | 3,7 mJ | 5,7 mJ | 7,5 mJ | 8,5 mJ |

Fig. 9. Expected extractable energy depending on the number of temporal replicas for the described experimental parameters, i.e. the fiber and stretched pulse duration employed in [45], based on the determined damage threshold of the fiber.

According to calculations (Fourier Split-Step approach with the amplification modeled by the Frantz-Nodvik equation [59]) using the parameters of the system used in [45] together with the experimentally determined damage threshold of the fiber (900 µJ), the total extractable energy depending on the number of pulse replicas can be extrapolated as shown in Fig. 9. As can be seen, by increasing the number of pulse replicas to 16 or even 32, the best part of the energy stored in the fiber could be extracted. However, as described in the experimental results, this will require good control of the pulse amplitudes and phases. Additionally, by increasing the stretched pulse duration, the number of required pulse replicas for efficient extraction could also be reduced.

## Conclusion

Coherent beam combination has matured from proof-of-principle experiments into a viable laser performance scaling method. Even though the approach can be applied to any laser architecture, it is especially well-suited for the combination of fiber laser systems due to their comparably simple setup and reproducible beam parameters. The resulting laser systems have already surpassed the performance of single amplifier systems and are now also being used for scientific applications. Many different beam combination and phase stabilization methods can be applied depending on the targeted parameter set. So far, most of the employed components have been adapted from single amplifier systems, but in order to obtain an even larger scalability, the development of integrated components and amplifiers will be required. The development, which has already begun, is expected to fulfil the promise that coherent combination can make physical limitations of single laser amplifiers mostly irrelevant for the total system performance. In the future, the final system performance will be determined by economic constraints solely and will enable new and exciting applications in a broad variety of scientific fields.

## References

[1] M. Kienel, M. Müller, S. Demmler, J. Rothhardt, A. Klenke, T. Eidam, J. Limpert, and A. Tünnermann, "Coherent beam combination of Yb:YAG single-crystal rod amplifiers.," *Opt. Lett.*, vol. 39, no. 11, pp. 3278–81, Jun. 2014.

[2] G. D. Goodno, H. Komine, S. J. McNaught, S. B. Weiss, S. Redmond, W. Long, R. Simpson, E. C. Cheung, D. Howland, P. Epp, M. Weber, M. McClellan, J. Sollee, and H. Injeyan, "Coherent combination of high-power, zigzag slab lasers," *Opt. Lett.*, vol. 31, no. 9, p. 1247, May 2006.

[3] S. M. Redmond, K. J. Creedon, J. E. Kansky, S. J. Augst, L. J. Missaggia, M. K. Connors, R. K. Huang, B. Chann, T. Y. Fan, G. W. Turner, and A. Sanchez-Rubio, "Active coherent beam combining of diode lasers.," *Opt. Lett.*, vol. 36, no. 6, pp. 999–1001, Mar. 2011.

[4] S. N. Bagayev, V. I. Trunov, E. V Pestryakov, S. A. Frolov, V. E. Leshchenko, A. E. Kokh, and V. A. Vasiliev, "Super-intense femtosecond multichannel laser system with coherent beam combining," *Laser Phys.*, vol. 24, no. 7, p. 74016, Jul. 2014.

[5] A. V Smith and J. J. Smith, "Mode instability in high power fiber amplifiers.," *Opt. Express*, vol. 19, no. 11, pp. 10180–92, May 2011.

[6] C. Jauregui, T. Eidam, H.-J. Otto, F. Stutzki, F. Jansen, J. Limpert, and A. Tünnermann, "Physical origin of mode instabilities in high-power fiber laser systems.," *Opt. Express*, vol. 20, no. 12, pp. 12912–25, Jun. 2012.

[7] C. Jauregui, J. Limpert, and A. Tünnermann, "High-power fibre lasers," *Nat. Photonics*, vol. 7, no. 11, pp. 861–867, Oct. 2013.

[8] T. Eidam, S. Hanf, E. Seise, T. V Andersen, T. Gabler, C. Wirth, T. Schreiber, J. Limpert, and A. Tünnermann, "Femtosecond fiber CPA system emitting 830 W average output power," *Opt. Lett.*, vol. 35, no. 2, pp. 94–96, 2010.

[9] G. Agrawal, *Nonlinear Fiber Optics*, 3rd ed. Academic Press, 2007.

[10] R. W. Boyd, *Nonlinear Optics*. Academic Press, Inc., 1992.

[11] D. Strickland and G. Mourou, "Compression of amplified chirped optical pulses," *Opt. Commun.*, vol. 56, no. 3, pp. 219–221, 1985.

[12] D. N. Schimpf, J. Limpert, and A. Tünnermann, "Optimization of high performance ultrafast fiber laser systems to >10 GW peak power," *J. Opt. Soc. Am. B*, vol. 27, no. 10, p. 2051, Sep. 2010.

[13] J. E. Ripper, "OPTICAL COUPLING OF ADJACENT STRIPE-GEOMETRY JUNCTION LASERS," *Appl. Phys. Lett.*, vol. 17, no. 9, p. 371, Oct. 1970.

[14] W. Chang, T. Zhou, L. A. Siiman, and A. Galvanauskas, "Femtosecond pulse spectral synthesis in coherently-spectrally combined multi-channel fiber chirped pulse amplifiers.," *Opt. Express*, vol. 21, no. 3, pp. 3897–910, Feb. 2013.

[15] A. Klenke, E. Seise, J. Limpert, and A. Tünnermann, "Basic considerations on coherent combining of ultrashort laser pulses,"

[16] L. Daniault, M. Hanna, D. N. Papadopoulos, Y. Zaouter, E. Mottay, F. Druon, and P. Georges, "Passive coherent beam combining of two femtosecond fiber chirped-pulse amplifiers.," *Opt. Lett.*, vol. 36, no. 20, pp. 4023–5, Oct. 2011.
[17] S. Szatmari and P. Simon, "Interferometric multiplexing scheme for excimer amplifiers," *Opt. Commun.*, vol. 98, no. 1–3, pp. 181–192, Apr. 1993.
[18] S. Zhou, F. W. Wise, and D. G. Ouzounov, "Divided-pulse amplification of ultrashort pulses," *Opt. Lett.*, vol. 32, no. 7, p. 871, 2007.
[19] H. Stark, M. Müller, M. Kienel, A. Klenke, J. Limpert, and A. Tünnermann, "Electro-optically controlled divided-pulse amplification," *Opt. Express*, vol. 25, no. 12, p. 13494, Jun. 2017.
[20] H. Tünnermann and A. Shirakawa, "Delay line coherent pulse stacking," *Opt. Lett.*, vol. 42, no. 23, p. 4829, Dec. 2017.
[21] M. Hanna, F. Guichard, Y. Zaouter, D. N. Papadopoulos, F. Druon, and P. Georges, "Coherent combination of ultrafast fiber amplifiers," *J. Phys. B At. Mol. Opt. Phys.*, vol. 49, no. 6, p. 62004, Mar. 2016.
[22] G. D. Goodno, C.-C. Shih, and J. E. Rothenberg, "Perturbative analysis of coherent combining efficiency with mismatched lasers.," *Opt. Express*, vol. 18, no. 24, pp. 25403–25414, 2010.
[23] L. Daniault, M. Hanna, L. Lombard, Y. Zaouter, E. Mottay, D. Goular, P. Bourdon, F. Druon, and P. Georges, "Impact of spectral phase mismatch on femtosecond coherent beam combining systems.," *Opt. Lett.*, vol. 37, no. 4, pp. 650–2, Feb. 2012.
[24] J. Bourderionnet, C. Bellanger, J. Primot, and A. Brignon, "Collective coherent phase combining of 64 fibers.," *Opt. Express*, vol. 19, no. 18, pp. 17053–8, Aug. 2011.
[25] R. Uberna, A. Bratcher, and B. G. Tiemann, "Coherent Polarization Beam Combination," *IEEE J. Quantum Electron.*, vol. 46, no. 8, pp. 1191–1196, Aug. 2010.
[26] E. C. Cheung, J. G. Ho, G. D. Goodno, R. R. Rice, J. Rothenberg, P. Thielen, M. Weber, and M. Wickham, "Diffractive-optics-based beam combination of a phase-locked fiber laser array," *Opt. Lett.*, vol. 33, no. 4, p. 354, Feb. 2008.
[27] P. a Thielen, J. G. Ho, D a Burchman, G. D. Goodno, J. E. Rothenberg, M. G. Wickham, A. Flores, C. a Lu, B. Pulford, C. Robin, A. D. Sanchez, D. Hult, and K. B. Rowland, "Two-dimensional diffractive coherent combining of 15 fiber amplifiers into a 600 W beam.," *Opt. Lett.*, vol. 37, no. 18, pp. 3741–3, 2012.
[28] T. Zhou, T. Sano, and R. Wilcox, "Coherent combination of ultrashort pulse beams using two diffractive optics," *Opt. Lett.*, vol. 42, no. 21, p. 4422, Nov. 2017.
[29] R. Harney and J. Schipper, *Passive and active pulse stacking scheme for pulse shaping*. United States: U.S. patent 4059759, 1977.
[30] H. Carstens, N. Lilienfein, S. Holzberger, C. Jocher, T. Eidam, J. Limpert, A. Tünnermann, J. Weitenberg, D. C. Yost, A. Alghamdi, Z. Alahmed, A. Azzeer, A. Apolonski, E. Fill, F. Krausz, and I. Pupeza, "Megawatt-scale average-power ultrashort pulses in an enhancement cavity.," *Opt. Lett.*, vol. 39, no. 9, pp. 2595–8, May 2014.
[31] T. W. Hansch and B. Couillaud, "Laser frequency stabilization by polarization spectroscopy of a reflecting reference cavity," *Opt. Commun.*, vol. 35, no. 3, pp. 441–444, Dec. 1980.
[32] T. M. Shay, "Theory of electronically phased coherent beam combination without a reference beam," *Opt. Express*, vol. 14, no. 25, p. 12188, 2006.
[33] M. A. Vorontsov, G. W. Carhart, and J. C. Ricklin, "Adaptive phase-distortion correction based on parallel gradient-descent optimization," *Opt. Lett.*, vol. 22, no. 12, p. 907, Jun. 1997.
[34] E. Seise, A. Klenke, S. Breitkopf, M. Plötner, J. Limpert, and A. Tünnermann, "Coherently combined fiber laser system delivering 120 μJ femtosecond pulses.," *Opt. Lett.*, vol. 36, no. 4, pp. 439–41, Feb. 2011.
[35] S. B. Weiss, M. E. Weber, and G. D. Goodno, "Group delay locking of coherently combined broadband lasers.," *Opt. Lett.*, vol. 37, no. 4, pp. 455–7, Feb. 2012.
[36] G. D. Goodno and S. B. Weiss, "Automated co-alignment of coherent fiber laser arrays via active phase-locking," *Opt. Express*, vol. 20, no. 14, p. 14945, Jun. 2012.
[37] M. Kienel, A. Klenke, T. Eidam, S. Hädrich, J. Limpert, and A. Tünnermann, "Energy scaling of femtosecond amplifiers using actively controlled divided-pulse amplification.," *Opt. Lett.*, vol. 39, no. 4, pp. 1049–52, Feb. 2014.
[38] T. Zhou, J. Ruppe, C. Zhu, I.-N. Hu, J. Nees, and A. Galvanauskas, "Coherent pulse stacking amplification using low-finesse Gires-Tournois interferometers," *Opt. Express*, vol. 23, no. 6, p. 7442, Mar. 2015.
[39] S. Breitkopf, T. Eidam, A. Klenke, L. von Grafenstein, H. Carstens, S. Holzberger, E. Fill, T. Schreiber, F. Krausz, A. Tünnermann, I. Pupeza, and J. Limpert, "A concept for multiterawatt fibre lasers based on coherent pulse stacking in passive cavities," *Light Sci. Appl.*, vol. 3, no. 10, p. e211, Oct. 2014.
[40] E. Seise, A. Klenke, J. Limpert, and A. Tünnermann, "Coherent addition of fiber-amplified ultrashort laser pulses.," *Opt. Express*, vol. 18, no. 26, pp. 27827–35, Dec. 2010.
[41] L. Daniault, M. Hanna, L. Lombard, Y. Zaouter, E. Mottay, D. Goular, P. Bourdon, F. Druon, and P. Georges, "Coherent beam combining of two femtosecond fiber chirped-pulse amplifiers.," *Opt. Lett.*, vol. 36, no. 5, pp. 621–3, Mar. 2011.
[42] J. Limpert, F. Stutzki, F. Jansen, H.-J. Otto, T. Eidam, C. Jauregui, and A. Tünnermann, "Yb-doped large-pitch fibres: effective single-mode operation based on higher-order mode delocalisation," *Light Sci. Appl.*, vol. 1, no. 4, p. e8, Apr. 2012.
[43] E. Treacy, "Optical pulse compression with diffraction gratings," *Quantum Electron. IEEE J.*, vol. 5, no. 9, pp. 454–458, Sep. 1969.
[44] M. Müller, M. Kienel, A. Klenke, T. Gottschall, E. Shestaev, M. Plötner, J. Limpert, and A. Tünnermann, "1 kW 1 mJ eight-channel ultrafast fiber laser," *Opt. Lett.*, vol. 41, no. 15, p. 3439, Aug. 2016.
[45] M. Kienel, M. Müller, A. Klenke, J. Limpert, and A. Tünnermann, "12 mJ kW-class ultrafast fiber laser system using multidimensional coherent pulse addition," *Opt. Lett.*, vol. 41, no. 14, p. 3343, Jul. 2016.
[46] M. Mueller, M. Kienel, A. Klenke, T. Eidam, J. Limpert, and A. Tünnermann, "Phase stabilization of spatiotemporally multiplexed ultrafast amplifiers," *Opt. Express*, vol. 24, no. 8, p. 7893, Apr. 2016.
[47] A. Ali, "On laser air breakdown, threshold power and laser generated channel length," *Nrl*, 1983.
[48] T. Gustafson, P. Kelly, and R. Fisher, "Subpicosecond pulse generation using the optical Kerr effect," *IEEE J. Quantum Electron.*, vol. 5, no. 6, pp. 325–325, Jun. 1969.
[49] S. Hädrich, A. Klenke, J. Rothhardt, M. Krebs, A. Hoffmann, O. Pronin, V. Pervak, J. Limpert, and A. Tünnermann, "High photon flux table-top coherent extreme-ultraviolet source," *Nat. Photonics*, vol. 8, no. 10, pp. 779–783, Sep. 2014.
[50] J. Rothhardt, S. Hädrich, Y. Shamir, M. Tschnernajew, R. Klas, A. Hoffmann, G. K. Tadesse, A. Klenke, T. Gottschall, T. Eidam, J. Limpert, A. Tünnermann, R. Boll, C. Bomme, H. Dachraoui, B. Erk, M. Di Fraia, D. A. Horke, T. Kierspel, T. Mullins, A. Przystawik, E. Savelyev, J. Wiese, T. Laarmann, J. Küpper, and D. Rolles, "High-repetition-rate and high-photon-flux 70 eV high-harmonic source for coincidence ion imaging of gas-phase molecules," *Opt. Express*, vol. 24, no. 16, p. 18133, Aug. 2016.
[51] G. K. Tadesse, R. Klas, S. Demmler, S. Hädrich, I. Wahyutama, M. Steinert, C. Spielmann, M. Zürch, T. Pertsch, A. Tünnermann, J. Limpert, and J. Rothhardt, "High speed and high resolution table-top nanoscale imaging," *Opt. Lett.*, vol. 41, no. 22, p. 5170, Nov. 2016.
[52] L. P. Ramirez, M. Hanna, G. Bouwmans, H. El Hamzaoui, M. Bouazaoui, D. Labat, K. Delplace, J. Pouysegur, F. Guichard, P. Rigaud, V. Kermène, A. Desfarges-Berthelemot, A. Barthélémy, F. Prévost, L. Lombard, Y. Zaouter, F. Druon, and P. Georges, "Coherent beam combining with an ultrafast multicore Yb-doped fiber amplifier," *Opt. Express*, vol. 23, no. 5, p. 5406, Mar. 2015.
[53] F. Prevost, L. Lombard, J. Primot, L. P. Ramirez, L. Bigot, G. Bouwmans, and M. Hanna, "Coherent beam combining of a narrow-linewidth long-pulse Er^3+-doped multicore fiber amplifier," *Opt. Express*, vol. 25, no. 9, p. 9528, May 2017.
[54] A. Klenke, M. Wojdyr, M. Müller, M. Kienel, T. Eidam, H.-J. Otto, F. Stutzki, F. Jansen, J. Limpert, and A. Tünnermann, "Large-pitch Multicore Fiber for Coherent Combination of Ultrashort Pulses," in *2015 European Conference on Lasers and Electro-Optics - European Quantum Electronics Conference*, 2015, p. CJ_1_2.
[55] H.-J. Otto, A. Klenke, C. Jauregui, F. Stutzki, J. Limpert, and A. Tünnermann, "Scaling the mode instability threshold with multicore fibers.," *Opt. Lett.*, vol. 39, no. 9, pp. 2680–3, May 2014.
[56] H.-J. Otto, F. Stutzki, N. Modsching, C. Jauregui, J. Limpert, and A.

Tünnermann, "2 kW average power from a pulsed Yb-doped rod-type fiber amplifier.," *Opt. Lett.*, vol. 39, no. 22, pp. 6446–9, Nov. 2014.
[57] T. Bifano, "Adaptive imaging: MEMS deformable mirrors," *Nat. Photonics*, vol. 5, no. 1, pp. 21–23, Jan. 2011.
[58] D. R. Herriott and H. J. Schulte, "Folded Optical Delay Lines," *Appl. Opt.*, vol. 4, no. 8, p. 883, Aug. 1965.
[59] L. M. Frantz and J. S. Nodvik, "Theory of Pulse Propagation in a Laser Amplifier," *J. Appl. Phys.*, vol. 34, no. 8, p. 2346, Jun. 1963.

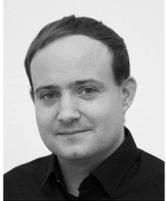

**Arno Klenke** was born in Paderborn, Germany, on September 24, 1984. He received the Diploma in general physics from the Friedrich-Schiller-Universität Jena, Germany, in 2010, followed by his Ph.D. in 2016.

Currently, he is working as a post-doc at the Helmholtz-Institute Jena. His research is focused on coherent combination of high power fiber lasers and ultrashort pulse amplifiers.

Mr. Klenke is member of the German Physical Society and the Optical Society of America.

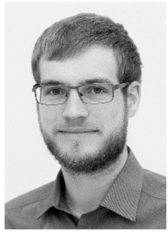

**Michael Mueller** was born in Heilbad Heiligenstadt, Germany, on December 2, 1990. He received his M.Sc. in Photonics from the Friedrich Schiller University Jena, Germany, in 2015.

Currently, he is working towards the Ph.D. degree at the Institute of Applied Physics, Friedrich-Schiller-University Jena. His research is focused on coherent combination of ultrashort pulses and optical frequency conversion.

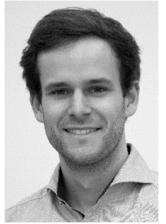

**Henning Stark** was born in Lübeck, Germany, on April 8, 1991. He received his B.Sc. in general physics from the Georg-August-Universität Göttingen, Germany, in 2014 and his M.Sc. in general physics from the Friedrich-Schiller-Universität Jena, Germany, in 2016.

Currently, he is working towards the Ph.D. degree at the Institute of Applied Physics in Jena. His primary research concerns are temporally separated amplification and spatio-temporal coherent combination of ultrashort pulses.

.

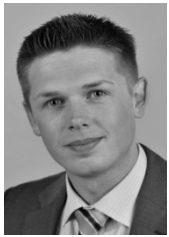

**Marco Kienel** was born in Meerane, Germany, on September 28, 1984. He received his M.Sc. in photonics from the Friedrich-Schiller-Universität Jena, Germany, in 2013.

He received his Ph.D. in 2017 and joined Active Fiber Systems GmbH thereafter, continuing his work on high-power fiber laser systems in industry.

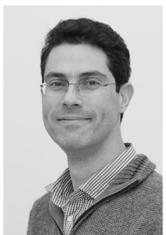

**Cesar Jauregui** was born in Santander, Spain, on June 18, 1975. He received both his Telecommunication Technical Engineering degree and his Telecommunication Engineering degree at the University of Cantabria. In 2003, he got his Ph.D. degree at that same University. In 2005 he began a two-year post-doc stay at the Optoelectronics Research Centre, Southampton, UK, where he investigated the phenomenon of slow-light in optical fibers.

Since 2007 he is working at the Institute of Applied Physics in Jena. His primary research concerns are high-power fiber lasers, non-linear effects and mode instabilities in optical fibers.

Dr. Jauregui is member of the Optical Society of America.

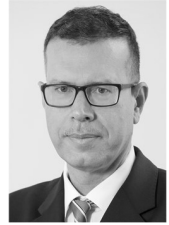

**Andreas Tünnermann** was born in Ahnsen, Germany, on June 10th, 1963. He received both his degree and Ph.D. degrees in Physics from the University of Hannover in 1988 and 1992, respectively. He was head of the department of development at the Laser Zentrum Hannover from 1992 to 1997. In the beginning of 1998 he joined the Friedrich-Schiller-University in Jena, Germany, as Director of the Institute of Applied Physics. In 2003 he became the Director of the Fraunhofer Institute of Applied Optics and Precision Engineering in Jena. He was appointed, in addition, as a Director of the Helmholtz-Institute Jena.

His main research interests revolve around scientific and technical aspects associated with the tailoring of light. He has authored and co-authored more than 500 peer-reviewed publications in international journals and has given more than 120 invited talks at national and international conferences. His research activities on optics and applied quantum electronics have been awarded with the Gottfried Wilhelm Leibniz Prize in 2005. He additionally received the Schott-award, the Leibinger-Innovation award, the Röntgen-prize and the WLT-prize. In 2011 he received the Thuringian order of merit. In 2015 his research activities have been awarded with an ERC advanced grant.

Prof. Tünnermann is a fellow of the Optical Society of America and SPIE.

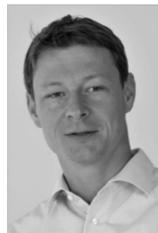

**Jens Limpert** was born in Jena, Germany, on 11th of December in 1975. He received his M.S. in 1999 and Ph.D. in Physics from the Friedrich-Schiller-Universität Jena, Germany, in 2003. After a one-year postdoc position at the University of Bordeaux, France, where he extended his research interests to high intensity lasers and nonlinear optics, he returned to Jena and is currently leading the Laser Development Group (including fiber- and waveguide lasers) at the Institute of Applied Physics. Together with his colleagues, he has invented novel large-mode-area fiber designs based on micro- and nanostructures. He has also developed novel experimental strategies and, based on these strategies, demonstrated significant power scaling of high repetition rate ultrafast lasers systems, as well as new concepts of optical parametric interaction and high-harmonic generation. He is author or co-author of more than 200 peer-reviewed journal papers in the field of laser physics. His research activities have been awarded with the WLT-Award in 2006, an ERC starting grant in 2009 and an ERC consolidator grant in 2013.

Prof. Limpert is member of the German Physical Society and the Optical Society of America.